\documentclass[12pt]{article}
\usepackage{graphicx}
\usepackage{graphics}
\usepackage{amssymb}
\usepackage{amscd}
\usepackage{feynmf}
\begin{document}
\begin{titlepage}

%\begin{center}
%{\hbox to\hsize{
%\hfill \bf hep-ph/??? }}
{\hbox to\hsize{\hfill September 2013 }}

\bigskip \vspace{3\baselineskip}

\begin{center}
{\bf \large 
Excluding a Generic Spin-2 Higgs Impostor }

\bigskip

\bigskip

{\bf Archil Kobakhidze and Jason Yue \\ }

\smallskip

{ \small \it
ARC Centre of Excellence for Particle Physics at the Terascale, \\
School of Physics, The University of Sydney, Sydney, NSW 2006, Australia.\\
E-mails: archil.kobakhidze@physics.usyd.edu.au; j.yue@physics.usyd.edu.au
\\}

\bigskip
 
\bigskip

\bigskip

{\large \bf Abstract}
\end{center}

\noindent 

We argue that the current experimental data in conjunction with perturbative unitarity considerations exclude the possibility that the LHC  $125-126$  GeV  resonance is a generic massive spin-2 particle of either parity. We analyse tree-level $Z-$spin-2 elastic scattering and demonstrate that perturbative unitarity breaks down at energies $\Lambda\sim 600$ GeV. Furthermore, we find that $W,Z-$spin-2 interactions contribute to the electroweak oblique parameters in a way that is in gross disagreement with observations.

\end{titlepage}
\baselineskip=16pt
\section{Introduction}

The bosonic resonance at  $m_h\approx 125-126$ GeV observed  by ATLAS \cite{Aad:2012tfa} and CMS \cite{Chatrchyan:2012ufa}  Collaborations has properties which closely resemble those of the Standard Model (SM) Higgs boson. The coupling  strengths measured from the rates in different decay channels are in good agreement with the SM predictions within the current experimental errors \cite{ATLAS:2013sla, CMS:yva}. The scalar (CP-even) nature of the observed resonance is strongly favoured over the pure CP-odd property \cite{Chatrchyan:2012jja, ATLAS:2013nma}, which is again in agreement with the SM expectation. Furthermore, the observation of the resonance in the diphoton channel excludes spin-1 particle due to the Landau-Yang theorem \cite{Yang:1950rg}, leaving a spin-2 particle as the only alternative to the spin-0 Higgs boson.  Due to the variety of different phenomenological models for spin-2 interactions,  discriminating between the spin-0 and spin-2 cases in full generality is quite a challenging task. The current studies are typically based on a graviton-like spin-2 model with minimal couplings. Accordingly, the ATLAS Collaboration excludes such spin-2 particle at 99.9\% C.L. based on their combined analyses of the $WW$, $ZZ$ and $\gamma\gamma$ decay channels, independent of the assumed contributions of gluon fusion and quark-antiquark annihilation processes in the production of the spin-2 particle \cite{ATLAS:2013mla}. The CMS Collaboration excludes the graviton-like spin-2 with minimal couplings at 99.4\% C.L. based on the combined analysis of the $WW$ and $ZZ$ channels, under the assumption that the spin-2 particle is produced in gluon fusion only \cite{CMS:yva}.  Although the above measurements are an important step forward in determining the spin and parity of the Higgs-like resonance, they are far from being fully conclusive,  since the possibility of a more generic spin-2 resonance of either parity still remains. Therefore, studies of various methods of spin discrimination  attracted a renewed interest after the LHC discovery \cite{Bolognesi:2012mm}.    
 
One of the main ingredients of the Higgs mechanism is that the scalar resonance ensures perturbative unitarity of high-energy scattering processes that involve massive weak gauge bosons. A tensor resonance cannot be a part of the Higgs mechanism and, furthermore, the perturbative unitarity can be further jeopardised by the spin-2 longitudinal states. The aim of this Letter is to analyse spin-2--weak gauge boson scattering in order to identify the energy scale where the theory enters a nonperturbative regime. Assuming that (non-universal) couplings of the spin-2 to the Standard Model gauge fields reproduce the experimentally observed production and decay rates, we calculate the amplitude for $hZ \to hZ$ elastic scattering and demonstrate that it grows as $\sim E^8$ with energy $E$, and, as a result, theory becomes strongly coupled at rather low energies, $E\approx 600$ GeV. Addition of allowed massive resonance(s) (e.g. spin-1,2 and 3) cannot improve this situation.  Furthermore,  we compute contributions $W,Z-$spin-2 interactions to the electroweak oblique parameters and find that they are in gross disagreement with observations.

This Letter is organised as follows: In Section 2 we discuss the theoretical shortcomings of a massive spin-2 field and set up the phenomenological model for our calculations. The calculation of  $hZ \to hZ$ amplitude is presented in Section 3, followed by Section 4 where the electroweak oblique parameters are discussed. Concluding remarks and further discussion are given in the final Section 5.

\section{Massive spin-2} 

Constructing a consistent theory of interacting massive spin-2 field, $h_{\mu\nu}(x)$, is a rather challenging theoretical task \cite{ArkaniHamed:2002sp} (see also \cite{Hinterbichler:2011tt} for a recent review). Such an effective theory  must in some way (e.g., non-linear realisation) incorporate  local gauge symmetry under the following transformation:
\begin{equation}
\delta_{\xi}h_{\mu\nu}=\partial_{\mu}\xi_{\nu}+\partial_{\nu}\xi_{\mu}
\label{1}
\end{equation}
where $\xi_{\mu}(x)$ is an arbitrary vector field which parameterises the infinitesimal local gauge transformations. This gauge invariance describes redundancies in the description and must be imposed to render negative norm states unphysical. 

In the linearised approximation, the Lagrangian of a non-interacting spin-2 field must take the Fierz-Pauli form \cite{Fierz:1939ix}:
\begin{eqnarray}
\label{2}
{\cal L}_{\rm FP}={\cal E}-\frac{1}{4}m_h^2\left(h_{\mu\nu}h^{\mu\nu}-h^2\right)~, \\
\label{3}
{\cal E}=\frac{1}{4}\partial_{\rho}h_{\mu\nu}\partial^{\rho}h^{\mu\nu}-\frac{1}{2}\partial_{\mu}h^{\mu\rho}\partial^{\nu}h_{\nu\rho} +\frac{1}{2}\partial_{\mu}h^{\mu\nu}\partial_{\nu}h-\frac{1}{4}\partial_{\rho}h\partial^{\rho}h~,
\end{eqnarray}
where $h:= h^{\mu}_{\mu}$. 
The specific form of the mass (\ref{2}) and kinetic terms (\ref{3}) are dictated by the important on-shell condition:
\begin{equation}
\partial^{\mu}\left(h_{\mu\nu}-\eta_{\mu\nu}h\right)=0~. 
\label{4}
\end{equation}
Any other arrangement of the kinetic and mass terms would lead to a theory with one or more extra ghost states \cite{VanNieuwenhuizen:1973fi}. In fact, the above theory can be viewed as a gauged fixed version ($A_{\mu}=0$) of the manifestly gauge invariant theory, 
\begin{equation}
{\cal L}={\cal L}_{\rm FP}-\frac{1}{4}F_{\mu\nu}F^{\mu\nu}-m_h\left(h_{\mu\nu}\partial^{\mu}A^{\nu}-h\partial_{\mu}A^{\mu}\right),
\label{5}
\end{equation}
where $F_{\mu\nu}:=\partial_{\mu}A_{\nu}-\partial_{\nu}A_{\mu}$. The corresponding gauge invariant transformations are given by (\ref{1}) and:
\begin{equation}
\delta A_{\mu}=-\frac{1}{m_h}\xi_{\mu}~.
\label{6}
\end{equation}
The integrability conditions (\ref{4})  can then be used to remove one scalar and three vector ghost states. 
   
In the linear approximation, spin-2 interactions with other fields are written as $\varpropto h_{\mu\nu}T^{\mu\nu}$\footnote{At the end of the Letter we briefly discuss more general interactions of spin-2 with the Standard Model fields.}. The condition (\ref{4}) is still satisfied, provided that the energy-momentum tensor $T^{\mu\nu}$ of fields is divergenceless on-shell, $\partial_{\mu}T^{\mu\nu}=0$. This actually requires non-linear interaction terms for a spin-2 field, since the total conserved energy-momentum tensor depends on $h_{\mu\nu}(x)$.   Thus,  $\partial_{\mu}T^{\mu\nu}\approx0$ in the linearised approximation. One must then carefully examine whether for a given physical process the linearised approximation breaks down, and whether the theory enters a nonperturbative regime.  Also, the exact condition $\partial_{\mu}T^{\mu\nu}=0$ strictly implies that the spin-2 couples to all other fields with a universal coupling. 

Here we follow a phenomenological approach and disregard possible spin-2 self-interactions. We also assume that the couplings to Standard Model fields are not universal, i.e.:
\begin{equation}
{\cal L}_{\rm int}=-\frac{\kappa_i}{2}h^{\mu\nu}T^{i}_{\mu\nu}~,
\label{7}
\end{equation}
where index $i$ runs over Standard Model particle species. These assumptions are made to keep the model flexible enough to accommodate various experimental constraints. Indeed, to reproduce the LHC data, spin-2 resonance must couple to the Standard Model fields with a strength not too different from the weak interaction strength. 

As discussed above, such a resonance universally coupled to the Standard Model fields through $\kappa_i=\kappa,~\forall i$,  is excluded by experiments. A theoretical argument against the graviton-like spin-2 with minimal couplings has been provided in an earlier paper \cite{Ellis:2012mj} based on the analysis of production and decay rates of the spin-2 particle. Furthermore, the spin-2 self-interactions would create more serious troubles for the theory. Thorough discussion of the problems of a non-linear theory of massive spin-2 field can be found in \cite{ArkaniHamed:2002sp, Hinterbichler:2011tt}. 

In this Letter we accept (\ref{2}) and (\ref{7}) as an effective theory for a spin-2 resonance which is valid at energies below certain scale $\Lambda$. By identifying this resonance with the one observed at LHC, we determine $\Lambda$ by analysing the violation of perturbative unitarity in processes involving a massive spin-2 particle, such as elastic scatterings of massive spin-2 and massive weak vector bosons. New physics must then enter around $\Lambda$ to provide a completion (at least partial) of the effective theory.  At high energies, the scattering processes are dominated by longitudinal polarisations of spin-2 and weak vector bosons and are most sensitive to the strong coupling scale $\Lambda$. This motivates us to consider the $hZ \to hZ$ scattering processes. 

\section{Perturbative unitarity in $hZ \to hZ$ scattering}            

One of the well-established decay channels of the LHC resonance is the decay into pair of electroweak gauge bosons $WW^{*}$ and $ZZ^{*}$. Kinematically one of the gauge bosons in the pair must be off-shell. The observed event rates in $WW^{*}$ and $ZZ^{*}$ channels are in good agreement with the Standard Model predictions. Moreover, no significant deviation from the Standard Model couplings has been found so far in the data \cite{ATLAS:2013sla, CMS:yva} (see also the analysis in \cite{Falkowski:2013dza}).
Therefore, it is reasonable to assume that spin-2 couples to the Standard Model particles in such a way that the  
rates of various processes, in particular, $h\to gg$ and $h\to WW^{*}/ZZ^{*}$ coincide with those of the Standard Model Higgs boson\footnote{Since the ATLAS and CMS searches are optimised for the spin-0 case, one actually must take into account their efficiencies for the case of spin-2. For $h\rightarrow ZZ^{*}\rightarrow 4l$ event selection one uses only the individual lepton momenta, and hence we may assume that the spin-2 and spin-0 efficiencies are the same, that is, $\Gamma_{\rm spin-2}(h\rightarrow ZZ^*)=\Gamma_{\rm Higgs}(h\rightarrow ZZ^*)$. Search strategies for  $h\rightarrow WW^{*}\rightarrow 2l2\nu$, on the other hand, were shown to be $\simeq 1.9$ times less efficient for the spin-2 \cite{Ellis:2012jv}, hence, $\Gamma_{\rm spin-2}(h\rightarrow WW^*)\simeq 1.9 \Gamma_{\rm Higgs}(h\rightarrow WW^*)$. This correction have no significant effect on our final results and we will neglect it in what follows. We would like to thank an anonymous referee for bringing this point to our attention.}. 

The off-shell decay widths for Higgs boson and a spin-2 particle, into electroweak bosons, $V=W,Z$, are computed in \cite{Keung:1984hn} and \cite{Geng:2012hy}, respectively: 
\begin{eqnarray} 
\Gamma_{\rm Higgs}(h\rightarrow VV^*)&=&\frac{3g^4m_Hm_V^4}{512\pi^3 m_W^4}\delta_VF_{\rm Higgs}(\epsilon_V)~,\\
\Gamma_{\rm spin-2}(h\rightarrow VV^*)&=& 
\frac{g^2m_h^3\kappa_V^2m_V^2}{5120\pi^3m_W^2}\delta_V F_{\rm spin-2}(\epsilon_V)~,
\label{8}
\end{eqnarray}
where $\epsilon_V=m_V/m_h$, $\delta_W=1$, $\delta_Z=
\frac{1}{12}(7-\frac{40}{3}\sin^2\theta_W+\frac{160}{9}\sin^4\theta_W)$ and
the functions $F_{\rm Higgs}(\epsilon)$ and $F_{\rm spin-2}(\epsilon)$ are defined as follows:
\begin{eqnarray}
F_{\rm Higgs}(\epsilon)=\frac{3(20\epsilon^4-8\epsilon^2+1)}{(4\epsilon^2-1)^{1/2}} 
\cos^{-1} \left( \frac{3\epsilon^2-1}{2\epsilon^3}  \right)\nonumber \\ 
-(1-\epsilon^2)\left(\frac{47}{2} \epsilon^{2}-\frac{13}{2} +\frac{1}{\epsilon^2}\right) 
-3(4\epsilon^4-6\epsilon^2+1)\ln(\epsilon)~, 
\label{9}
\end{eqnarray}
\begin{eqnarray}
F_{\rm spin-2}(\epsilon)=\frac{368\epsilon^6+104\epsilon^4+29\epsilon^2-12}{(4\epsilon^2-1)^{1/2}}
\cos^{-1} \left( \frac{3\epsilon^2-1}{2\epsilon^3}  \right) \nonumber \\
-\frac{1}{60}(21\epsilon^{10}-200\epsilon^8-9150\epsilon^6+4560\epsilon^4+2765\epsilon^2+2004)
-(90\epsilon^6-30\epsilon^4+5\epsilon^2-12)\ln(\epsilon)~.
\label{10}
\end{eqnarray}
From the conditions $\Gamma_{\rm Higgs}(h\rightarrow VV^*)= 
\Gamma_{\rm spin-2}(h\rightarrow VV^*)$, we extract the couplings of spin-2 to $W$ and $Z$ bosons:
\begin{eqnarray}
\label{11a} 
  \kappa_W^2\approx 3.61 \cdot 10^{-5}~{\rm GeV}^{-2}~,  \\
\label{11b}  
  \quad  \kappa_Z^2\approx4.42 \cdot 10^{-5}~{\rm GeV}^{-2}~, 
  \end{eqnarray}
where we have used $m_h=125$ GeV, $m_Z= 91.2$ GeV, $m_W=80.4$ GeV and $\alpha_W=g^2/4\pi=1/29.58$ in the above numerical estimations. As is expected, $\kappa^2_{Z,W}$ are of the order of the weak Fermi constant $G_F\approx 1.17\times 10^{-5}$ GeV$^{-2}$. 

%FIGURE-1%%%%%%%%%%%%%%%%%%%%%%%%%%%%

\begin{figure}[t]
%\centering
\includegraphics[width=0.5\textwidth]{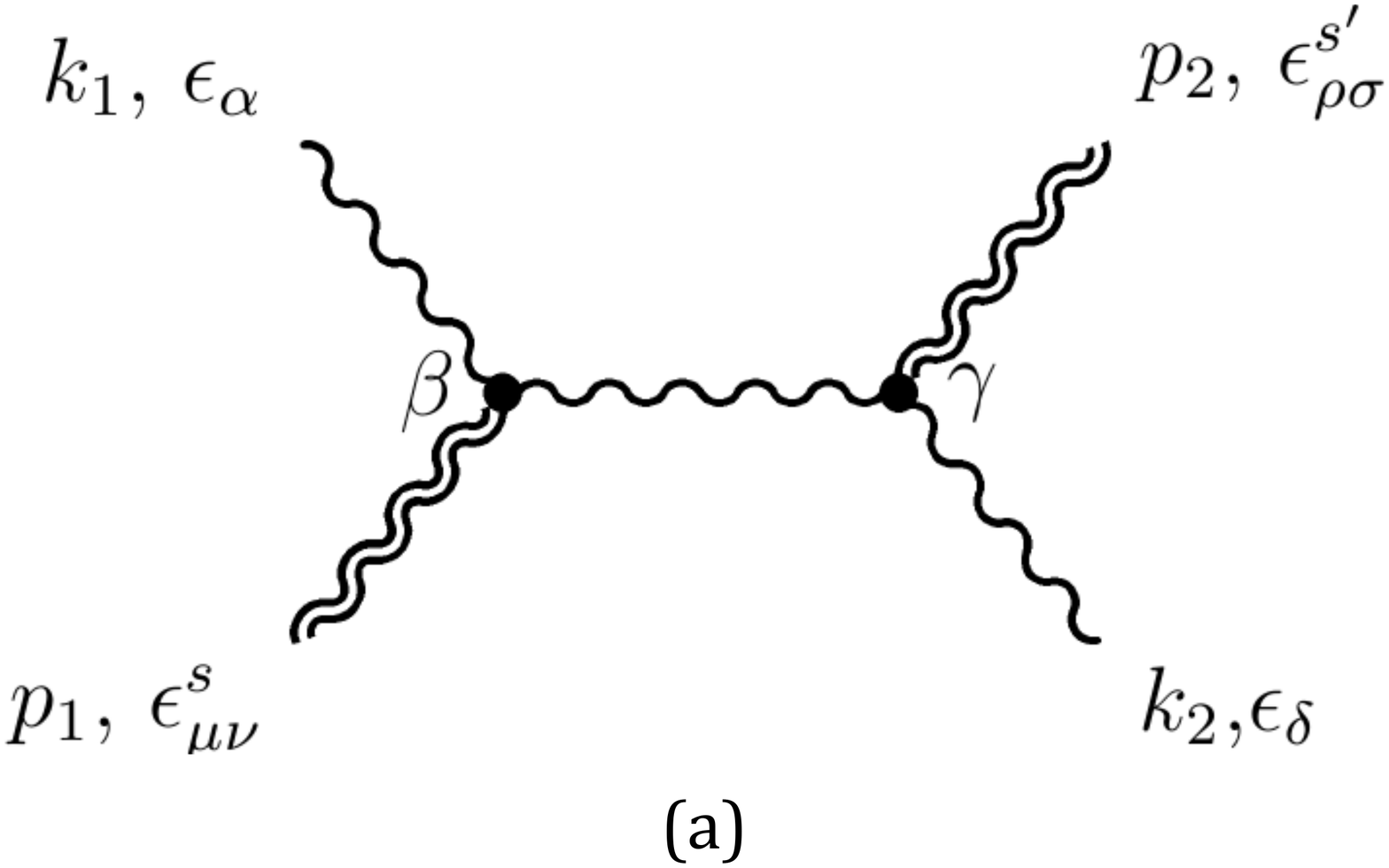}
\includegraphics[width=0.5\textwidth]{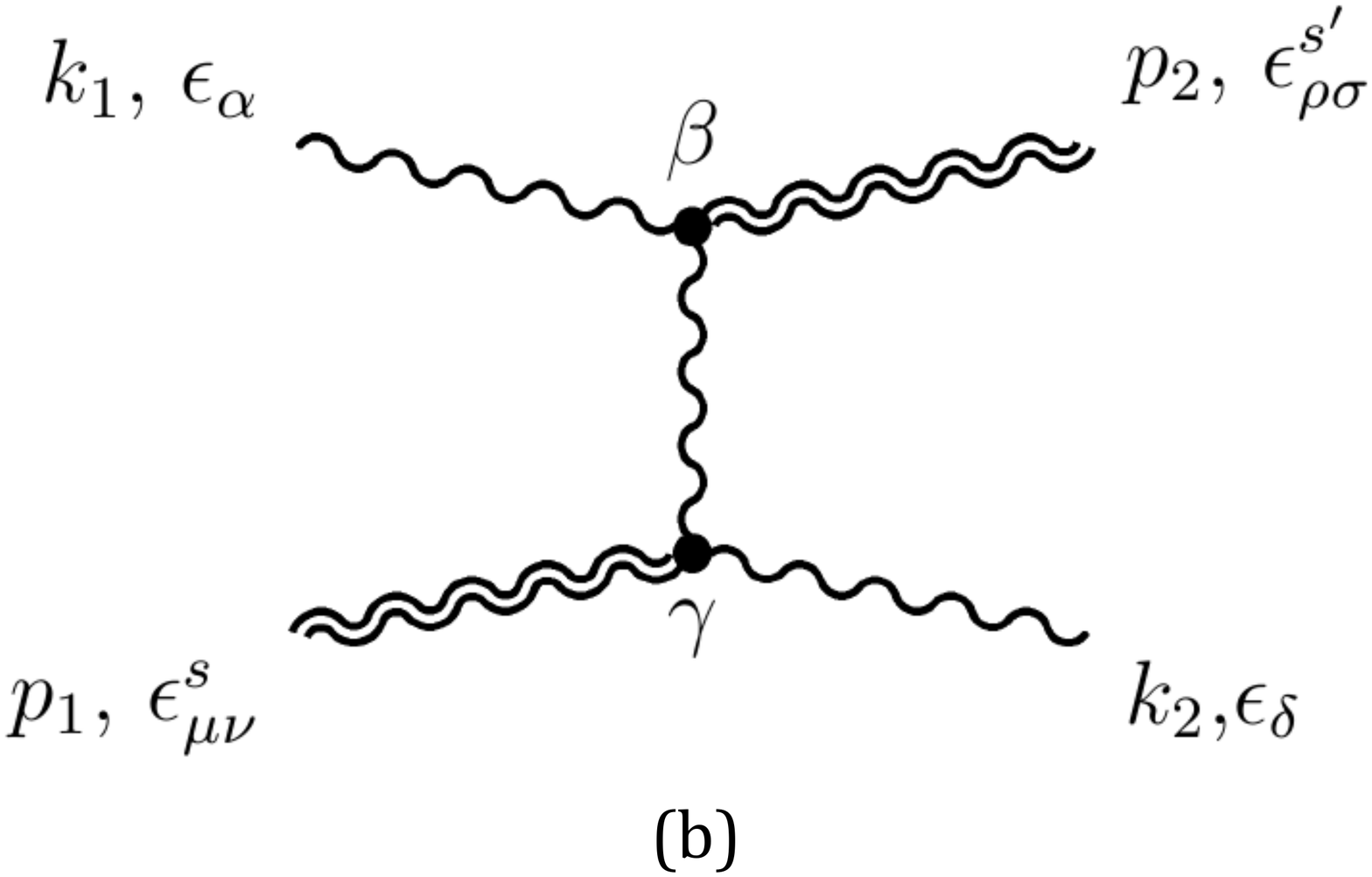}
\caption{ (a) s-channel and (b) t-channel Feynman diagrams for the tree-level $hZ\to hZ$ scattering. }
\label{fig1}
\end{figure}

%%%%%%%%%%%%%%%%%%%%%%%%%%%%%%%%%%%%

Next, we consider $hZ\to hZ$ elastic scattering. Using the appropriate Feynman rules given in e.g., \cite{Han:1998sg}, it is easy to establish that at the lowest level of perturbation theory there are two contributing Feynman diagrams (see Figure \ref{fig1}). At high energies, the longitudinal states dominate the scattering process. Since the longitudinal polarisation vector for the $Z$ boson and longitudinal scalar polarisation tensor for the spin-2 contain the highest powers in momentum, we concentrate on the scattering of these states. Naively, one expects that amplitude to grow as $\sim E^{10}$ with energy:  $E^1$ comes from each of the longitudinal external $Z$, $E^2$ comes from each of the longitudinal $h$ and each vertex, while longitudinal part of the $Z$ propagator contributes $E^0$.  However, contraction of the longitudinal propagator with vertex is proportional to $\sim m_Z$, and the amplitude actually grows as $\sim E^8$. In the high-energy limit, $s\gg m_h^2$, we find that the overall scattering amplitude for Figure \ref{fig1} takes the form $\mathcal{M}=\mathcal{M}_{a}+\mathcal{M}_{b}$, with:
\begin{eqnarray}
\mathcal{M}_{a} \approx -\frac{\kappa_Z^2 s^3}{24m_h^4}\left(1-4\frac{m_Z^2}{s}(1-\cos\theta)\right)~,
\label{12}
\end{eqnarray}
\begin{eqnarray}
\mathcal{M}_{b} \approx 
\frac{\kappa_Z^2s^4}{512m_h^4m_Z^2(1-\cos\theta)}\left(\csc^2 \frac{\theta}{2} \sin^6\theta + 8\frac{m_Z^2}{s}(1+\cos\theta)^4 \right)~,
\label{13}
\end{eqnarray} 
where $s=(p_1+k_1)^2$ and $\theta$ is the scattering angle. Expanding the amplitude in partial waves:
\begin{equation}
\mathcal{M}=32\pi\sum_{J=0}^{\infty}(2J+1)a_JP_J(\cos\theta)~,
\label{14}
\end{equation}
with $P_J(\cos\theta)$ being the Legendre polynomials, we find that the unitarity condition, $|\mathrm{Re}\, a_0|<1/2$, breaks down at rather low energies, $\sqrt{s}\approx 600$ GeV. Thus, we adopt an upper limit on the cut-off for our effective theory, 
\begin{equation}
\Lambda \approx 600~ {\rm GeV}~,
\label{15}
\label{equation} 
\end{equation} 
 above which the model for 125 GeV spin-2 Higgs impostor is not valid without further adjustments.
 
 %FIGURE-2%%%%%%%%%%%%%%%%%%%%%%%%%%%%
 
\begin{figure}[t]
%\centering
\includegraphics[width=0.5\textwidth]{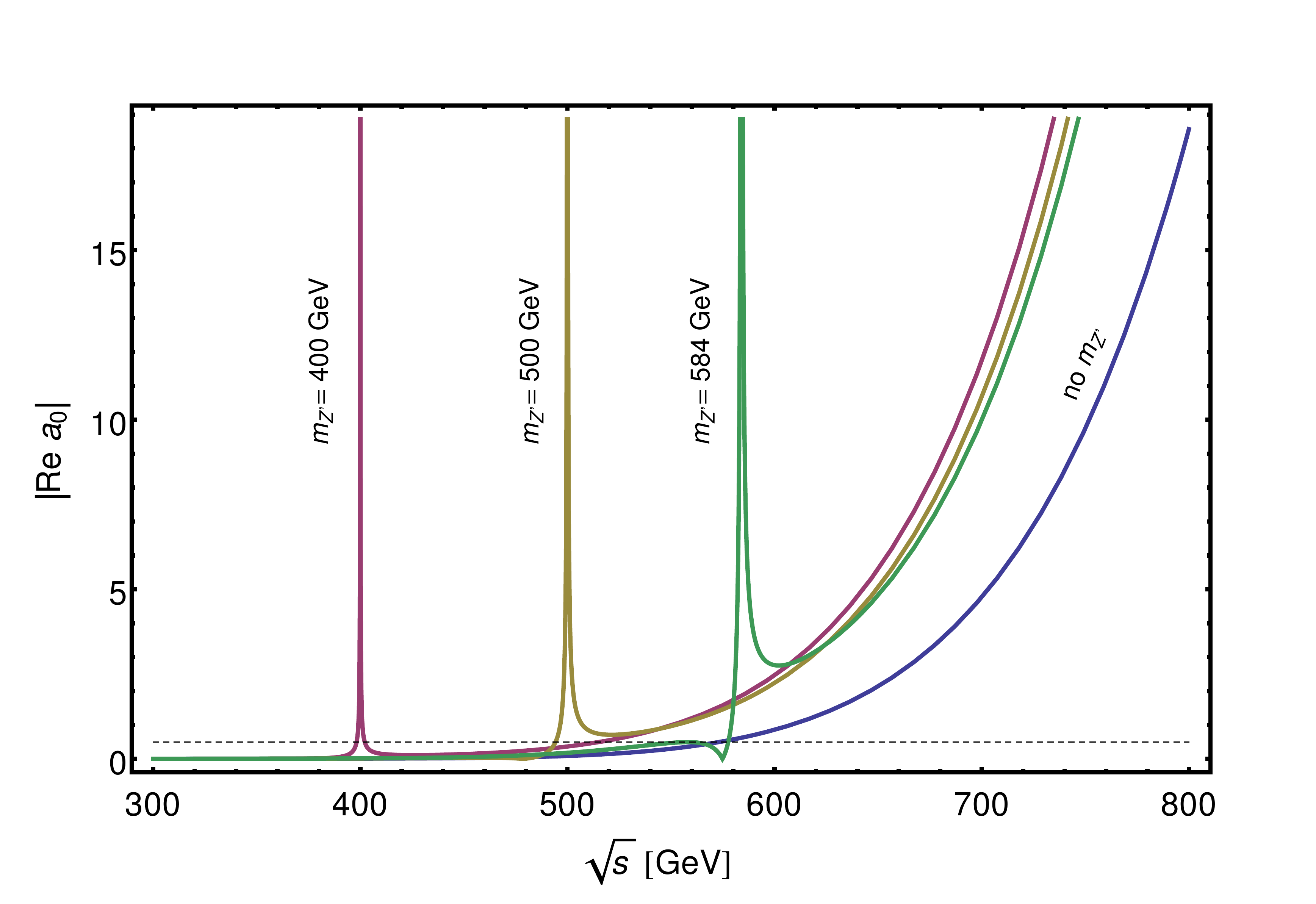}
\includegraphics[width=0.5\textwidth]{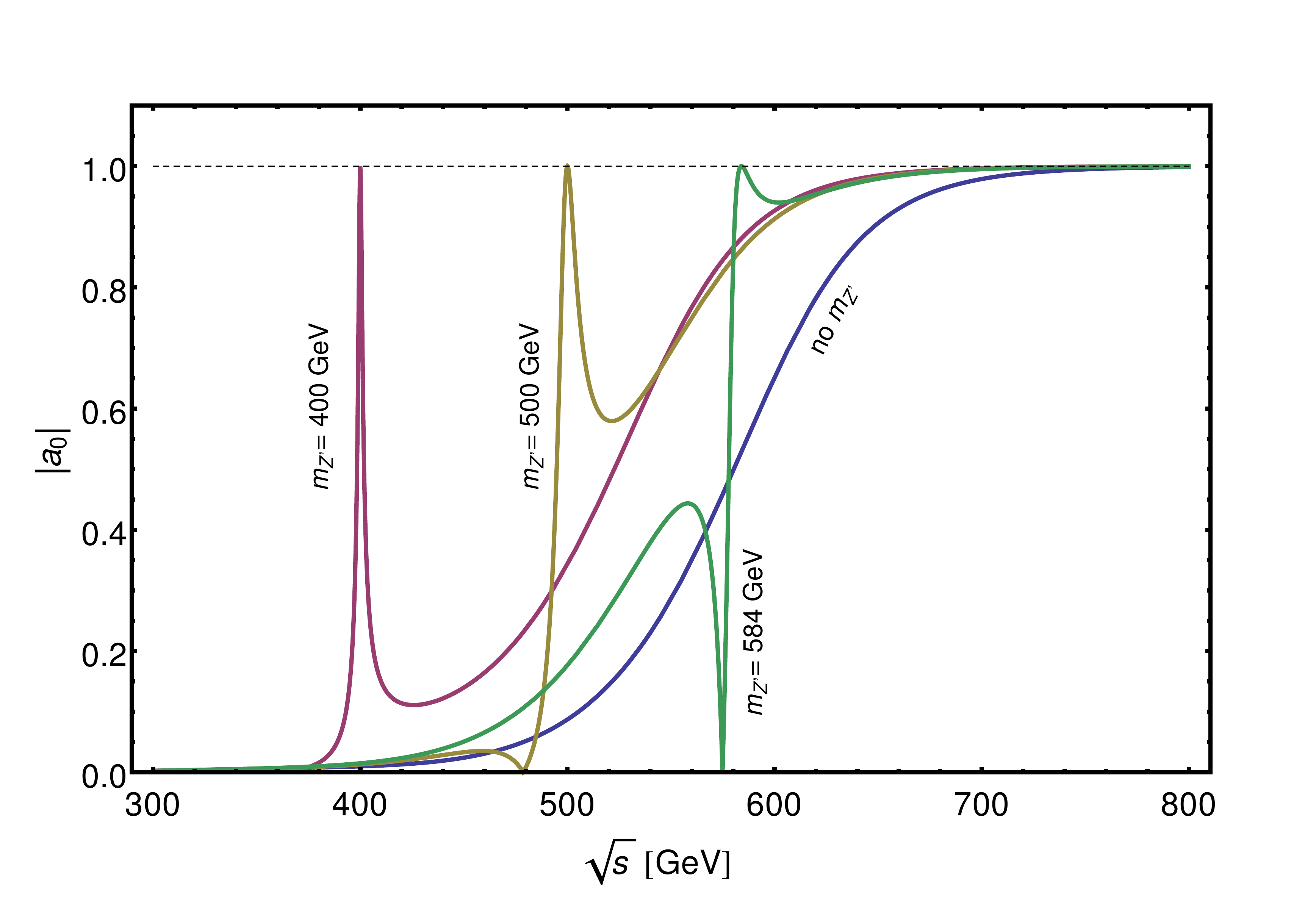} 
\\
\includegraphics[width=0.5\textwidth]{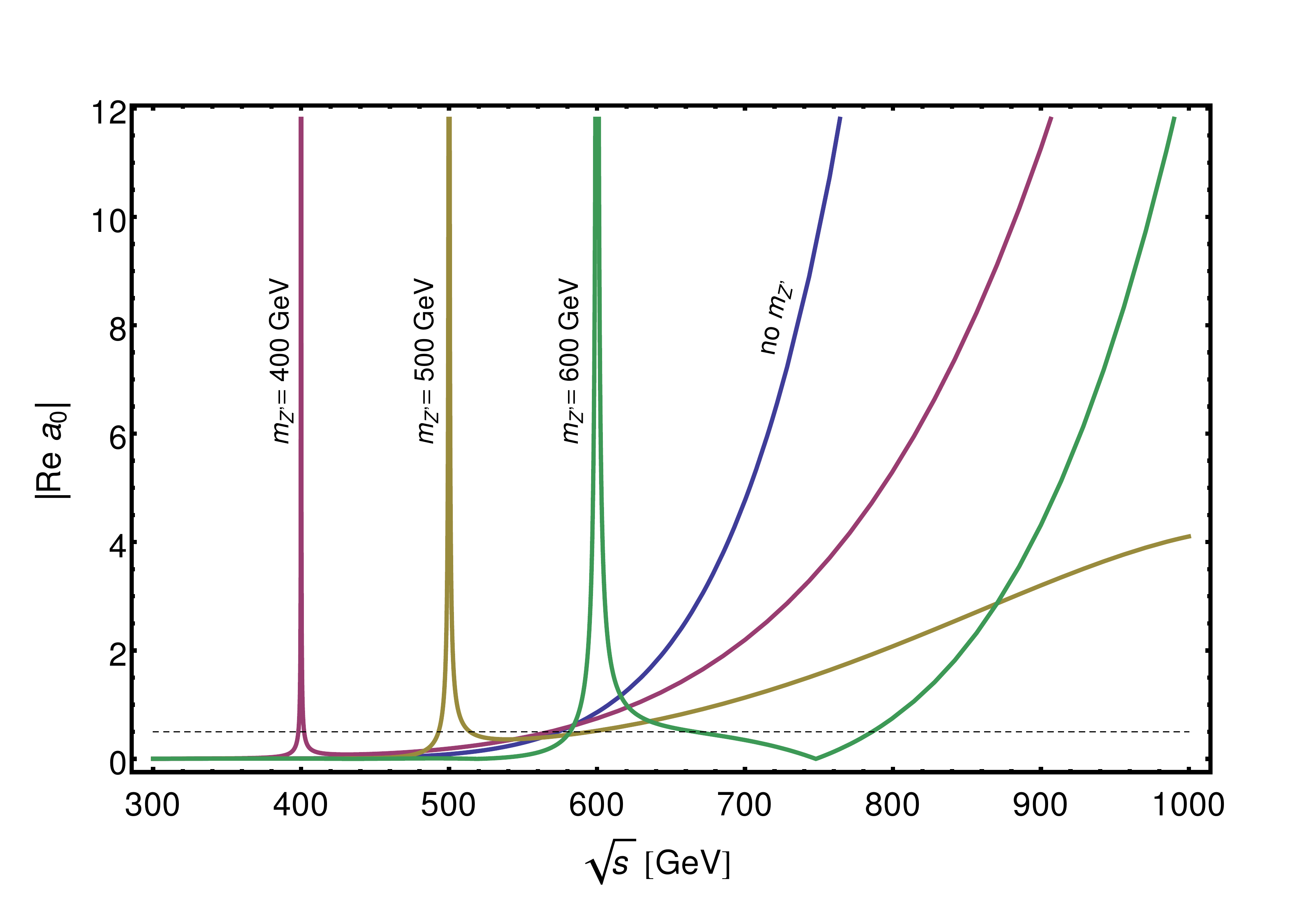}
\includegraphics[width=0.5\textwidth]{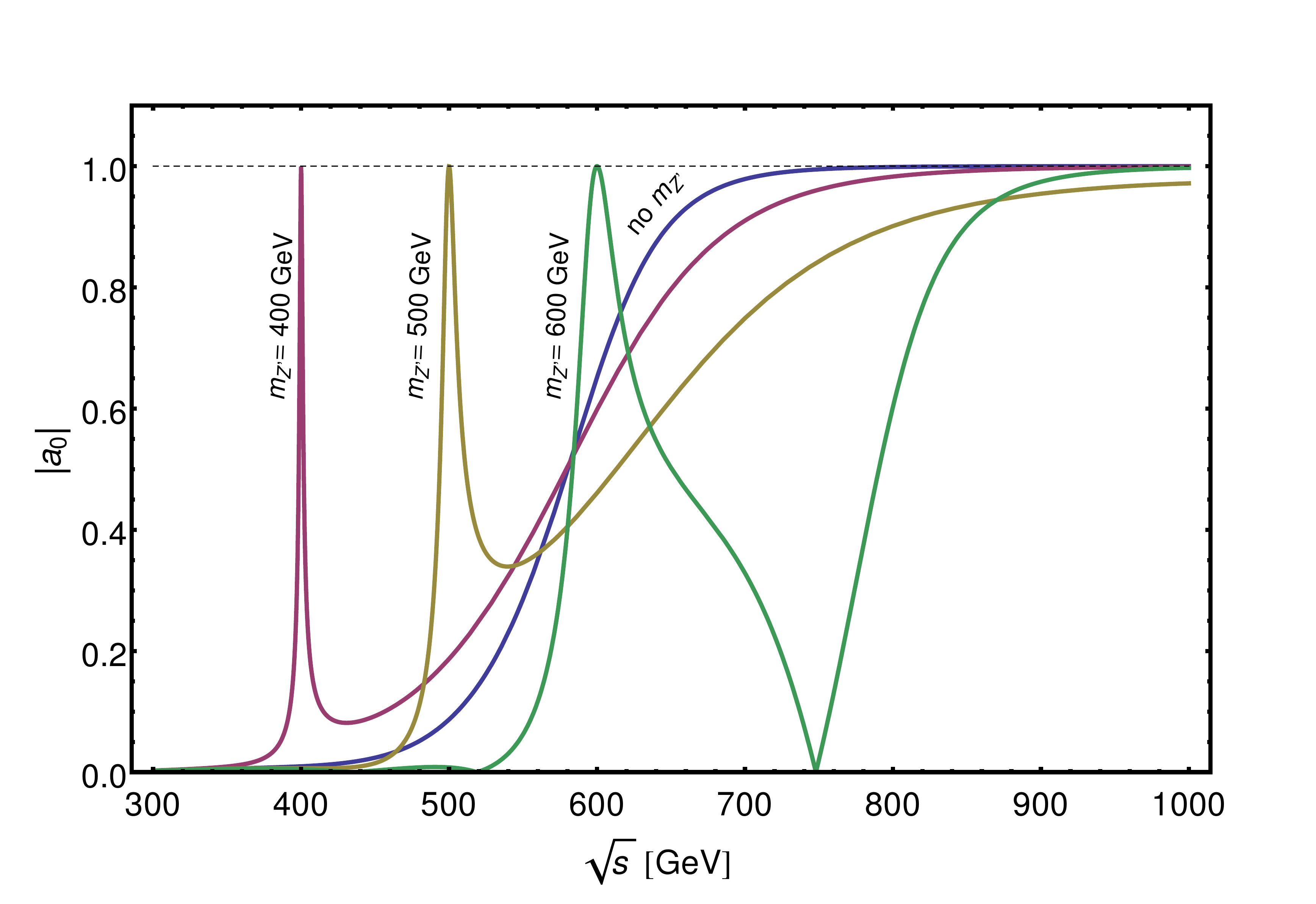}
\caption{Energy dependence of $|\rm{Re}\, a_0|$ of a non-unitarised (top left panel) partial wave amplitude $a_0$ within an effective theory with 125 GeV spin-2 impostor and an extra $Z'$ resonance of mass $m_{Z'}$ ($\kappa_{Z'}=\kappa_{Z}$ is assumed).  Corresponding unitarised amplitudes within the K-matrix formalism are shown on the top right panel. The same is depicted on the lower panels for $\kappa_{Z'}=i\kappa_{Z}$, so that the leading $\sim s^4$ energy dependence is cancelled out. }
\label{fig2}
\end{figure} 

%%%%%%%%%%%%%%%%%%%%%%%%%%%%%%%%%%%%%
 
 One may wonder whether the above situation can be improved by introducing an extra resonance, similar to way the Higgs boson improves the high-energy behaviour of scattering amplitudes involving weak gauge bosons. Recall that in, e.g., $WW\to WW$ scattering, the leading energy dependence $\sim E^4$ is automatically cancelled due to the (non-linearly realised) gauge invariance, while cancellation of the $\sim E^2$ dependence is due to the specific Higgs couplings to the electroweak gauge bosons dictated by the linearly realised spontaneously broken  gauge invariance. However,  no such invariance is at our disposal, and no similar cancellations are expected, as is confirm by the direct calculations. Possible resonances in the $hZ$-channel are those with spin-1, 2 and 3. We disregard spin-2 and 3 because scattering amplitudes of these states with spin-2 generally contain higher powers of energy, resulting in more severe violation of unitarity. Extra spin-1 resonance $Z'$ may have $hZZ'$ coupling of the sort (\ref{7}), e.g., emerging through the electroweak gauge invariant kinetic mixing of $Z'$ with hypercharge gauge field. The coupling constant of this interactions are not constrained by any symmetry to cancel out the power-law dependence of scattering amplitudes in (\ref{12}) and (\ref{13}). As a result, such extra resonances are not capable of postponing breakdown of perturbative unitarity significantly (see, the left upper panel on  Figure \ref{fig2}), even if the couplings are fine tuned in such a way that the leading $\sim E^8$ dependence is cancelled out (see the left lower panel on  Figure \ref{fig2}). In the later case, the cut-off $\Lambda$ can be as large as $\sim 800$ GeV. We also employ $K-$matrix formalism \cite{Chung:1995dx}  to unitarise the above amplitudes. As can be seen on the right panels in Figure \ref{fig2}, the unitarised partial wave amplitude quickly approaches its limiting value at $\sim 700$ GeV for untuned couplings (the upper-right panel) and $\sim 900$ GeV for tuned couplings (the lower-right panel). That is to say, the effective theory requires a completion at energies $\lesssim 700-900$ GeV.   
 
 %FIGURE-3%%%%%%%%%%%%%%%%%%%%%%%%%%%%
         
\begin{figure}[t]
\centering
\vspace{-1cm}
\includegraphics[width=12cm,height=7.5cm]{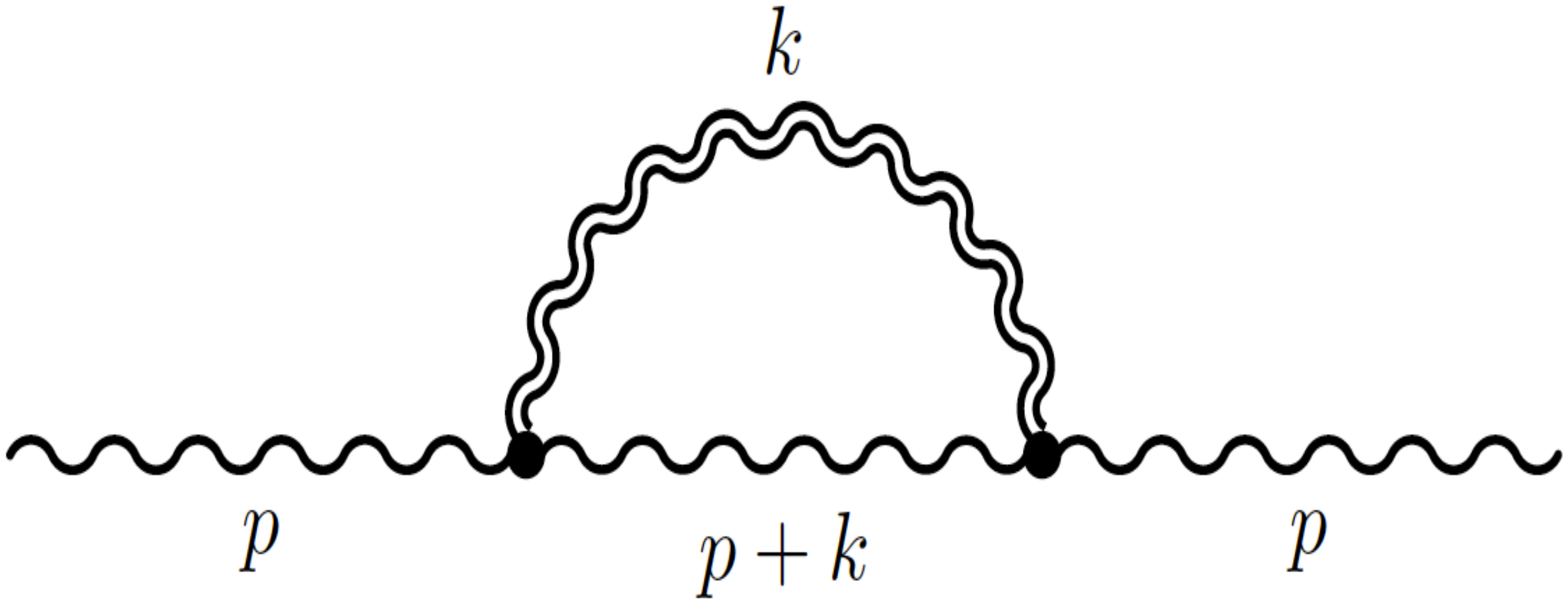}
\vspace{-2cm}
\caption{One-loop self-energy diagram  for the electroweak gauge bosons $V=W, Z$ due to the spin-2 impostor $h$.}
\label{fig3}
\end{figure} 

%%%%%%%%%%%%%%%%%%%%%%%%%%%%%%%%%%%%%

\section{Oblique corrections}

Let us compute now contribution of $hW$ and $hZ$ interactions to the electroweak oblique parameters \cite{Peskin:1991sw}: 
\begin{eqnarray}
   \alpha S &=&4s_{\theta_W}^2 c_{\theta_W}^2 \frac{\Pi_Z(m_Z^2)-\Pi_Z(0)}{m_Z^2}~,    \\
    \alpha T &=& \frac{\Pi_W(0)}{m_W^2}-\frac{\Pi_Z(0)}{m_Z^2}~,\\
    \alpha (S+U)&=& 4s_{\theta_W}^2 \frac{\Pi_W(m_W^2)-\Pi_W(0)}{m_W^2}~,
\label{16} 
\end{eqnarray}
where $\alpha\approx 1/127.91$ is the $\overline{\rm MS}$ running fine structure constant and $s_{\theta_W}^2=1-c_{\theta_W}^2\approx 0.2312$  is the sine squared of the weak mixing angle measured at $m_Z$. In the 1-loop approximation, contributions to the self-energies $\Pi_V$ in the above equations are given by the diagram depicted in Figure \ref{fig3}, with external legs being either $W$ or $Z$ bosons. The corresponding couplings are determined as in Eqs. (\ref{11a}) and  (\ref{11b}). Similar calculations in the case of Kaluza-Klein (KK) gravitons have been already performed in \cite{Han:2000gp}. We straightforwardly modified their calculations of the rainbow diagram by `removing' the summation over the massive KK states to give:   
\begin{eqnarray}
\Pi_V(m_V^2)&=&\frac{\kappa_V^2m_V^2 }{192\pi^2m_h^{4}}\int^{\Lambda^2}_0 dk_E^2 \int^1_0 dz\frac{k_E^2}{[k_E^2+m_h^2(1-z)+m_V^2z^2]^2}f_1(k_E^2,m_h^2,z,m_V^2)~,
\label{17}\\
\Pi_V(0)&=&\frac{\kappa^2 m_V^2}{192\pi^2m_h^{4}}\int^{\Lambda^2}_0 dk_E^2 \frac{k_E^2}{(k_E^2+m_V^2)(k_E^2+m_h^2)}f_2(k_E^2,m_h^2,m_V^2)~,
\end{eqnarray}
where 
\begin{eqnarray}
f_1(x,y,z,r) = 4r^3(z-2)^2z^4+r^2z^2\bigg[16y(z-2)-x(21z^2-52z+24)\bigg] \nonumber \\
+r\bigg[4xy(z^2-8z+2)+4y^2(9z^2+4)+x^2(15z^2-14z+1)\bigg]-x(x^2+4xy+23y^2)~,\\
f_2(x,y,r)  =x(x+y)(x+13y)+r(4x^2+26xy+52y^2)~.
\label{18}
\end{eqnarray}

%FIGURE-4%%%%%%%%%%%%%%%%%%%%%%%%%%%%

\begin{figure}[t]
\centering
\includegraphics[scale=0.4]{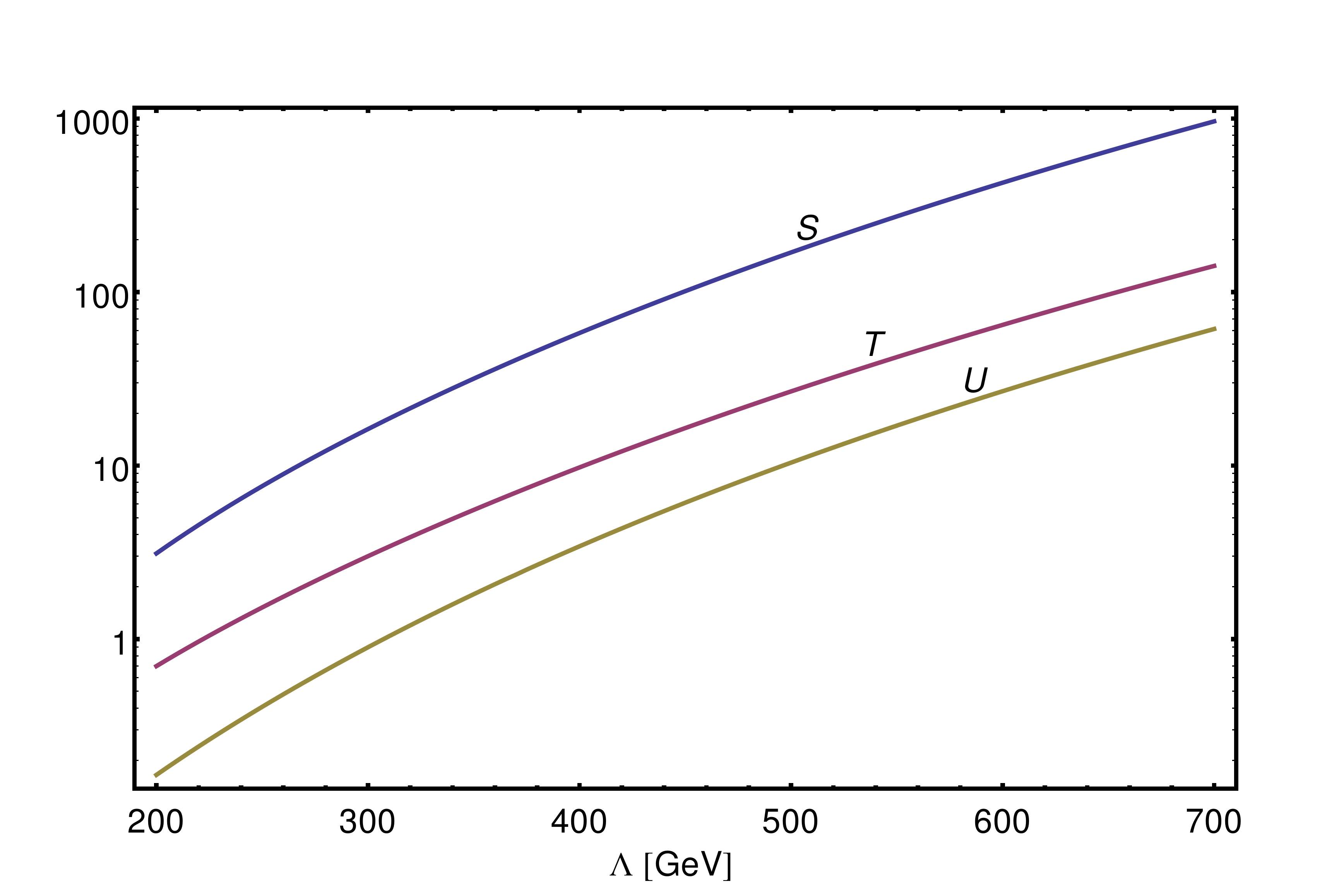}
\caption{$S$, $T$ and $U$ parameters as a function of cut-off scale $\Lambda$ in the effective theory with a 125 GeV spin-2 Higgs impostor. }
\label{fig4}
\end{figure} 

%%%%%%%%%%%%%%%%%%%%%%%%%%%%%%%%%%%%

%TABLE-1%%%%%%%%%%%%%%%%%%%%%%%%%%%%%

\begin{table}[t]
\begin{center}
{\renewcommand{\arraystretch}{1.25}
\begin{tabular}{ c  c  c  c  }
\hline 
{Parameter/Energy cut-off ($\Lambda$)} & 200 GeV  &400 GeV & 600 GeV         \\
\hline  
$S$ &-3.12&-58.0&-427.0 \\
$T$ & -0.70  &-9.75&-64.6\\
$U$& -0.17 &-3.42&-26.80\\
\hline
\end{tabular}}
\caption{Representative values of $S$, $T$ and $U$ parameters in the spin-2 Higgs impostor effective theory.} 
\vspace{-5ex}
\end{center}
\label{tab1}
\end{table}

%%%%%%%%%%%%%%%%%%%%%%%%%%%%%%%%%%%%%%

Our numerical results for $STU$ parameters are shown in Figure \ref{fig4} as functions of the cut-off scale $\Lambda$ and some representative values are presented in Table \ref{tab1}.  These predictions must be compared against experimental values \cite{Baak:2012kk}: 
\begin{eqnarray}
\label{19}
S= 0.03\pm 0.01~,\\
\label{20}
T=0.05\pm 0.12~,\\
\label{21}
U=0.03\pm 0.10~,
\end{eqnarray} 
which have however, been obtained by fitting electroweak observables for fixed reference values  $m_t^{\rm ref}=173$ GeV, $m_h^{\rm ref}=126$ GeV for top-quark and Higgs boson masses respectively. To remove the Higgs boson contribution, we correct the $S$ and $T$ parameters in Eqs. (\ref{20}) and  (\ref{21}) by
\begin{eqnarray}
\label{22}
\Delta S\approx \frac{1}{6\pi}\ln \left(\frac{\Lambda}{m_h^{\rm ref}}\right)~,\\
\label{23}
\Delta T= -\frac{3}{8\pi c^2_{\theta_W}}\ln\left(\frac{\Lambda}{m_h^{\rm ref}}\right)~,
\end{eqnarray}     
but $\Delta U=0$, as the  $U$ parameter is insensitive to the Higgs boson mass at the 1-loop level. We note immediately that the spin-2 Higgs impostor contribution to the $U$ parameter is compatible only marginally and only for very low cut-off $\Lambda \approx 200$ GeV with measured values in (\ref{21}), while predictions for the $S$ and $T$ parameters are in gross disagreement with experimental values (\ref{19}) and (\ref{20}). Furthermore, the corrections (\ref{22}) and (\ref{23}) have opposite signs and thus are not capable of removing this disagreement.

\section{Conclusion and discussion}
 
In this Letter we have considered the hypothesis that the 125 GeV LHC resonance is a spin-2 Higgs impostor with non-universal couplings to the Standard Model gauge bosons, and mimics the Standard Model Higgs production and decay rates. If true, this would imply that perturbative unitarity is broken at rather low energies, $\Lambda\approx 600$ GeV, and we cannot think of any sensible completion of the effective theory which is capable of postponing the breakdown of the theory to substantially higher energies.  Furthermore, we have demonstrated that the contributions of the $W,Z-$spin-2 interactions to the electroweak oblique parameters are in gross disagreement with the experimental data.

The conclusion that an effective theory with a spin-2 impostor breakdowns at rather low energies can readily be extended to interactions beyond those in Eq. (\ref{7}) and to the case of a spin-2 impostor with odd parity. Indeed, a quick inspection of generic interactions (see, e.g., \cite{Miller:2001bi}) reveals that for a spin-2, even-parity impostor, there are $hZZ$ vertices depending on 0, 2 and 4 momenta, while for an odd-parity impostor, each vertex contains at least two momenta (e.g., $~\epsilon^{\mu\nu\alpha\beta}p^{\rho}k_{\beta}$, etc.). Hence, the spin-2, even-parity amplitude grows at least as $\sim E^6$ and that of the odd parity at least as $\sim E^{8}$. Therefore, we expect that perturbative unitarity will be violated at energies $\Lambda \lesssim 1$ TeV.      
      
\paragraph{Acknowledgment} This work was supported by the Australian Research Council.

\end{document}